\documentstyle[12pt,epsf,epsfig]{article}
\let\section=\subsection     \let\subsection=\subsubsection                %%
\begin{document}
\begin{flushright}
SUNY-NTG-97-05
\end{flushright}
\begin{center}
   {\large \bf LOW-MASS e$^+$e$^-$ PAIRS FROM IN-MEDIUM }\\[2mm]
   {\large \bf $\rho$ MESON PROPAGATION}\\[5mm]
   R.~RAPP$^1$, G.~CHANFRAY$^2$ and J.~WAMBACH$^3$ \\[5mm]
   {\small \it  1) Department of Physics, SUNY at Stony Brook, 
     Stony Brook, NY 11794-3800, U.S.A. \\
    2) IPN-Lyon, 43 Av. de 11 Novembre 1918, F-69622 Villeurbanne Cedex,
    France \\
    3) Institut f\"ur Kernphysik, TH Darmstadt, Schlo{\ss}gartenstr.9,
    D-64289 Darmstadt, Germany 
    \\[8mm] }
\end{center}
%%%%%%%%%%%%%%%%%%%%%%%%%%%%%%%%%%%%%%%%%%%%%%%%%%%%%%%%%%%%%%%%%%%%%%%%%%
\begin{abstract}\noindent
Based on a realistic model for the rho meson in free space we investigate 
it's medium modifications in a hot hadron gas generated by hadronic 
rescattering processes, 
{\it i.e.} renormalization of intermediate two-pion states as well as 
direct rho meson scattering off hadrons. Within the vector dominance 
model the resulting in-medium rho spectral function is applied to calculate 
$e^+e^-$ spectra as recently measured in heavy-ion collisions at 
CERN-SpS energies in the CERES experiment.    
\end{abstract}
%%%%%%%%%%%%%%%%%%%%%%%%%%%%%%%%%%%%%%%%%%%%%%%%%%%%%%%%%%%%%%%%%%%%%%%%%%%
\section{Introduction}
%%%%%%%%%%%%%%%%%%%%%%%%%%%%%%%%%%%%%%%%%%%%%%%%%%%%%%%%%%%%%%%%%%%%%%%%%%
The main goal of ultrarelativistic heavy-ion collisions is the identification 
of possible phase transitions in strongly interacting matter associated with 
chiral symmetry restoration and/or deconfinement. Experimental signatures of 
such transitions have to be disentangled from hadronic rescattering processes 
occurring in the later stages of central collisions.  \\      
Even though electromagnetic probes (photons and dileptons) can traverse the 
hadronic interaction zone without further distortion, the eventually observed 
spectra will be contaminated with contributions arising from conventional hadronic 
mechanisms and even decays after the hadronic freezeout. Focussing on 
the dilepton production, various 'conventional' background radiation, depending 
on the invariant mass range, is to be expected: for $M_{l^+l^-}\ge$1.5~GeV    
($l$=$\mu,e$) Drell-Yan processes have to be disentangled from, {\it e.g.}, 
a possible enhancement due to thermal $q\bar q$ annihilation in a QGP, or 
from anomalous $J/\Psi$, $\Psi'$ suppression, which may or may not be due 
to a QGP formation; for $M_{l^+l^-}\le$1.5~GeV, the spectrum should be dominated 
by hadron decays. Here, the light vector mesons $\rho(770)$, $\omega(782)$ and 
$\phi(1020)$ are of particular interest, since they can directly couple to 
dilepton pairs. Among these the rho meson is of special importance due to it's 
short lifetime ($\tau_\rho^{free}$=1.3~fm/c), which is about an order of magnitude 
smaller than the typical lifetime of the hadronic fireball, 
$\tau_{fireball}$$\approx$10~fm/c (to be compared with $\tau_\omega^{free}$=23~fm/c
and $\tau_\phi^{free}$=44~fm/c). \\ 
A systematic study of low-mass $e^+e^-$ production in p-Be, p-Au and S-Au 
collisions at CERN-SpS energies has recently been performed by the CERES/NA45 
collaboration~\cite{CERES}. Whereas their event generator (accounting for 
'primary' hadron decays) can succesfully describe the p-A data, an overall 
factor of about 5 enhancement was observed in the S-Au case, reaching a 
maximum factor of $\sim$10 
around invariant masses $M_{e^+e^-}$$\simeq$0.4~GeV (similar results have 
been obtained by the HELIOS-3 collaboration~\cite{HELIOS3}; preliminary data 
from Pb-Au collisions further confirm these findings~\cite{Ulrich}). The inclusion 
of free $\pi^+\pi^-\to e^+e^-$ annihilation in transport~\cite{Cass,LKB} 
or hydrodynamical~\cite{Ming,Prak} simulations of the collision dynamics has 
been shown to reduce this discrepancy, still leaving a factor of up to 3 
too little yield below the $\rho$ mass. So far, the only {\it quantitative}
explanation of these data could be achieved by assuming a density and temperature 
dependent dropping $\rho$ mass according to the Brown-Rho scaling 
conjecture~\cite{BR91}, interpreted as a signature of (partial) chiral symmetry 
restoration. However, in this contribution we try to demonstrate that 
'conventional' hadronic rescattering mechanisms of the $\rho$ meson in a 
hot and dense hadronic enviroment seem to be sufficient to account for the 
experimentally observed $e^+e^-$ excess in central S-Au (200~GeV/u) and 
Pb-Au (158~GeV/u) collisions~\cite{RCW}.    
%%%%%%%%%%%%%%%%%%%%%%%%%%%%%%%%%%%%%%%%%%%%%%%%%%%%%%%%%%%%%%%%%%%%%%%%%%%%
\section{The Free $\rho$ Meson and Modifications in a Hot Hadron Gas}
%%%%%%%%%%%%%%%%%%%%%%%%%%%%%%%%%%%%%%%%%%%%%%%%%%%%%%%%%%%%%%%%%%%%%%%%%%%%
\subsection{The $\rho$ Meson in Free Space}
%%%%%%%%%%%%%%%%%%%%%%%%%%%%%%%%%%%%%%%%%%%%%%%%%%%%%%%%%%%%%%%%%%%%%%%%%%%
A satisfactory yet simple description of the $\rho$ meson in the vacuum can 
be achieved by renormalizing a 'bare' $\rho$ of mass $m_\rho^{bare}$ through 
coupling to intermediate two-pion states. The scalar part of the free $\rho$
propagator then reads  
\begin{equation}
D_\rho^0(M)=[M^2-(m_\rho^{bare})^2-\Sigma_{\rho\pi\pi}^0(M)]^{-1} \ ,
\label{drho0}
\end{equation}
where the selfenergy 
\begin{eqnarray}
\Sigma_{\rho\pi\pi}^0(M) & = & \bar{\Sigma}_{\rho\pi\pi}^0(M)
-\bar{\Sigma}_{\rho\pi\pi}^0(0) \ ,
 \nonumber\\
\bar{\Sigma}_{\rho\pi\pi}^0(M) & = & \int \frac{k^2 dk}{(2\pi)^2} \
v_{\rho\pi\pi}(k)^2 \ G_{\pi\pi}^0(M,k)  \  ,
\label{sigrho0}
\end{eqnarray}
contains the $\rho\pi\pi$ vertex function $v_{\rho\pi\pi}$ as well as 
the free two-pion propagator $G_{\pi\pi}^0(M,k)$. The subtraction at zero 
energy ensures the correct normalization of the pion electromagnetic 
form factor ($F_\pi(0)$=1), which in the vector dominance model (VDM) 
is given by
\begin{eqnarray}
|F_\pi^0(M)|^2 =  (m_\rho^{bare})^4 \ |D_\rho^0(M)|^2 \ .
\end{eqnarray}
The bare mass $m_\rho^{bare}$ and coupling $g_{\rho\pi\pi}$ (entering 
 $v_{\rho\pi\pi}$) are easily tuned to reproduce the experimental data 
on p-wave $\pi\pi$ scattering and the pion
electromagnetic form factor in the timelike region~\cite{RCW,CRW}. 
%%%%%%%%%%%%%%%%%%%%%%%%%%%%%%%%%%%%%%%%%%%%%%%%%%%%%%%%%%%%%%%%%%%%%%%%%%%%%
\subsection{Medium Modifications in $\pi\pi$ Propagation}  
%%%%%%%%%%%%%%%%%%%%%%%%%%%%%%%%%%%%%%%%%%%%%%%%%%%%%%%%%%%%%%%%%%%%%%%%%%%%%
The most important medium effects in the intermediate two-pion 
states of the $\rho$ propagator are attributed to interactions with 
surrounding baryons, as discussed in refs.~\cite{HeFN,AKLQ,ChSc} for the case 
of cold nuclear matter. Therefore one first needs a realistic model for the 
in-medium single-pion propagator $D_\pi$. As is well known from pion nuclear 
phenomenology, pion-induced p-wave nucleon-nucleonhole ($NN^{-1}$) and 
delta-nucleonhole ($\Delta N^{-1}$)
excitations are the dominant mechanism. Since we are interested in  
URHIC's at CERN-SpS energies (160-200~GeV/u), thermal excitations 
of the system should be taken into account, which, in the baryonic sector, 
are dominated by a large $\Delta$(1232) component. Thus we extend the 
particle-hole picture to include $\pi$-$\Delta$ interactions as well in 
form of $N\Delta^{-1}$ and $\Delta\Delta^{-1}$ excitations. \\  
To calculate the corresponding medium modified $\rho$ selfenergy, vertex 
corrections of the $\pi\pi\rho$ vertex have to be included to ensure 
the conservation of the vector current. We here employ the approach of 
Chanfray and Schuck~\cite{ChSc}. Within a full off-shell treatment of the 
pion propagation in connection with the afore mentioned extension to 
finite temperature the imaginary part of the in-medium $\rho$ selfenergy 
at zero 3-momentum can be cast in the form~\cite{CRW}  
\begin{eqnarray}
\lefteqn{ {\rm Im} \Sigma_{\rho\pi\pi}(q_0,\vec 0) = -
\int\limits_{0}^{\infty} \frac{k^2 \ dk}{(2\pi)^2}
\ v_{\rho\pi\pi}(k)^2 \int\limits_0^{q_0} \frac{dk_0}{\pi}
\ [1+f^\pi(k_0)+f^\pi(q_0-k_0)] } 
\nonumber\\
 & * & {\rm Im} D_\pi(q_0-k_0,k) \ {\rm Im}  \lbrace
\alpha(q_0,k_0,k) D_\pi(k_0,k)+\frac{\Pi_L(k_0,k)}{2k^2}+
\frac{\Pi_T(k_0,k)}{k^2} \rbrace .
\label{imsgrpp}
\end{eqnarray}
with the longitudinal and transverse spin-isospin response functions
$\Pi_L(k_0,k)$ and $ \Pi_T(k_0,k)$, a factor $\alpha$ characterizing 
vertex corrections and thermal Bose distributions $f^\pi$.    
The real part is obtained from a dispersion integral:
\begin{equation}
{\rm Re} \Sigma_{\rho\pi\pi}(q_0)=-{\cal P} \int\limits_0^\infty
\frac{dE'^2}{\pi} \frac{{\rm Im} \Sigma_{\rho\pi\pi}(E')}{q_0^2-E'^2}
\frac{q_0^2}{E'^2} \ .
\label{resgrpp}
\end{equation}
%%%%%%%%%%%%%%%%%%%%%%%%%%%%%%%%%%%%%%%%%%%%%%%%%%%%%%%%%%%%%%%%%%%%%%%%
\subsection{Rho-Nucleon and Rho-Delta Interactions} 
%%%%%%%%%%%%%%%%%%%%%%%%%%%%%%%%%%%%%%%%%%%%%%%%%%%%%%%%%%%%%%%%%%%%%%%%
In analogy to the pionic interactions with the surrounding medium 
direct interactions of the (bare) rho meson with nucleons 
and deltas may have a substantial impact. Based on the observation that 
certain baryonic excitations (especially $N$(1720) and $\Delta$(1905)) 
exhibit a strong coupling to the $\rho$$N$ 
decay channel (which suggests to identify them as '$\rho$$N$ resonances'), 
Friman and Pirner proposed to derive a corresponding in-medium $\rho$ 
selfenergy. As for pions, this is conveniently done in terms of 
p-wave ($\rho$-like) particle-hole excitations~\cite{FrPi}. The 
$\rho N(1720)N$ and $\rho\Delta(1905)N$ coupling constants are fixed 
by the experimental branching ratios (where it is very important 
to account for the finite $\rho$-width in free space to obtain realistic 
values).  Within our off-shell treatment we are also able to incorporate 
lower lying $\rho N$ and $\rho\Delta$ contributions, the coupling constants 
for which are taken from the Bonn potential. Thus we obtain 
\begin{equation}
\Sigma_{\rho B}(q_0,q)= -q^2 \ \sum_\alpha \chi_{\rho\alpha}(q_0,q)
\label{sgrbb}
\end{equation}
where the summation is over $\alpha=NN^{-1},\Delta N^{-1},
N\Delta^{-1},\Delta\Delta^{-1},N(1720)N^{-1}$, $\Delta(1905)N^{-1}$, and 
the susceptibilties $\chi_{\rho\alpha}$ contain the loop integrals of the 
corresponding particle-hole bubble as well as short-range correlation 
corrections (parametrized by Migdal paprameters $g'$)~\cite{RCW}. 
In fig.~1 we display the transverse part of the $\rho$ spectral function, 
\begin{equation}
{\rm Im} D_\rho^T(M,q)= {\rm Im} \left( \frac{1}{M^2-(m_\rho^{bare})^2
-\Sigma_{\rho\pi\pi}^0(M)-\Sigma_{\rho B}(q_0,q)} \right) \ ,
\end{equation}
at normal nuclear matter density and small temperature T=5~MeV with no  
medium modifications applied to the two-pion states. A pronounced 
structure of various branches is observed, in particular the 
$\rho N(1720)N^{-1}$, which may be phrased 'Rhosobar' (in analogy to the 
'Pisobar' $\pi\Delta N^{-1}$).  
\begin{figure}[h] 
\begin{center}
\epsfig{file=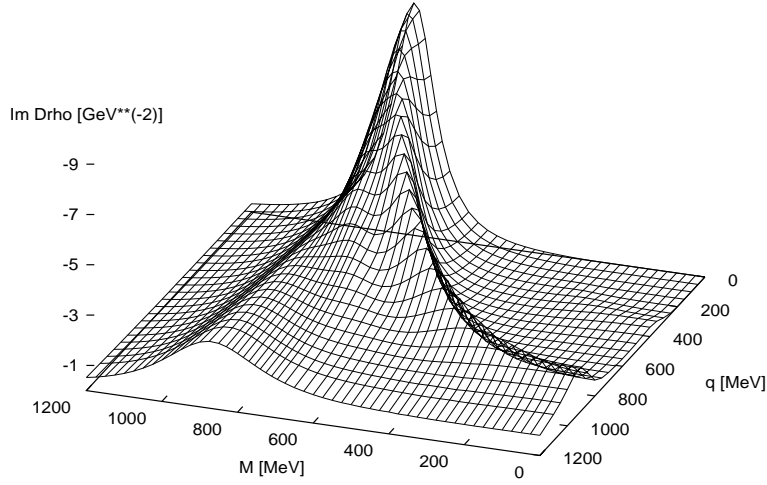,width=12cm,height=8cm, angle=-90}
\end{center}
\caption{\small{
Imaginary part of the $\rho$ propagator in cold nuclear matter
at a density $\rho$=$\rho_0$ when including 'Rhosobar-like' $\rho N$
interactions.} }
\end{figure}
%%%%%%%%%%%%%%%%%%%%%%%%%%%%%%%%%%%%%%%%%%%%%%%%%%%%%%%%%%%%%%%%%%%%%%%%%%%
\subsection{Rho-Pion and Rho-(Anti)Kaon Interactions}
%%%%%%%%%%%%%%%%%%%%%%%%%%%%%%%%%%%%%%%%%%%%%%%%%%%%%%%%%%%%%%%%%%%%%%%%%%%
Since in URHIC's at CERN-SpS energies large numbers of secondaries are 
produced, we furthermore evaluate $\rho$ scattering off the most abundant 
surrounding mesons, {\i.e.} pions and (anti-) kaons. Assuming the interactions 
to be dominated by $a_1$(1260) and $K_1$/$\bar{K_1}$(1270) formation, the 
corresponding $\rho$ selfenergy (in the Matsubara approach) can be written as 
\begin{equation}
\Sigma_{\rho M}(q_0,\vec q)=\int \frac{d^3p}{(2\pi)^3}
\frac{1}{2\omega_p^M} [f^M(\omega_p^M)-
f^{\rho M}(\omega_p^M+q_0)] \ M_{\rho M}(p_M,q)
\end{equation}
($M$=$\pi$, $K$, $\bar K$). The invariant scattering amplitude $M_{\rho M}$ is 
derived from a suitable (gauge invariant) lagrangian~\cite{XSB}. \\ 
As long as the meson chemical potentials are kept zero the effect of $\rho$-$M$
scattering is rather small: at highest temperatures considered ($T$=170~MeV) we 
find a $\sim$60~MeV broadening of the $\rho$ spectral function~\cite{RCW}, 
which is similar to the results of ref.~\cite{Hagl}.    
%%%%%%%%%%%%%%%%%%%%%%%%%%%%%%%%%%%%%%%%%%%%%%%%%%%%%%%%%%%%%%%%%%%%%%%%%% 
\section{$e^+e^-$ Spectra from in-Medium $\pi^+\pi^-$ Annihilation at the 
CERN-SpS} 
%%%%%%%%%%%%%%%%%%%%%%%%%%%%%%%%%%%%%%%%%%%%%%%%%%%%%%%%%%%%%%%%%%%%%%%%%%
Invoking the phenomenologically well established VDM the dilepton prodcution
rate from $\pi^+\pi^-$ annihilation can be expressed in terms of the $\rho$ 
spectral function as 
\begin{equation}
{dN_{\pi^+\pi^-\to e^+e^-}\over d^4xd^4q} =
-\frac{\alpha^2 (m_\rho^{bare})^4}{\pi^3 g_{\rho\pi\pi}^2} \
\frac{f^\rho(q_0;T)}{M^2} \ {\rm Im}D_\rho(q_0,q;\mu_B,T)
\label{rate}
\end{equation}
with 
\begin{equation}
{\rm Im}D_\rho  =  \frac{1}{3} \ \left( 
\frac{{\rm Im} \Sigma_\rho^{L}}{|M^2-(m_\rho^{bare})^2 -\Sigma_\rho^{L}|^2} 
+  \frac{2 {\rm Im} \Sigma_\rho^{T}}{|M^2-(m_\rho^{bare})^2 -
\Sigma_\rho^{T}|^2} 
\right)
\end{equation}
The full in-medium $\rho$ selfenergy $\Sigma_\rho$ is the sum of the 
contributions discussed in sects.~2.2-2.4 (decomposed in transverse and 
longitudinal parts)~\cite{RCW}. 
Fig.~2 shows the $\rho$ spectral function at fixed chemical potentials and given 
three-momentum: with increasing temperature/density a dramatic broadening is 
found, which, in particular, results in a pronounced enhancement over the free 
curve for invariant masses below $M$$\simeq$0.6~GeV.  
\begin{figure}[h]
\begin{center}
\epsfig{file=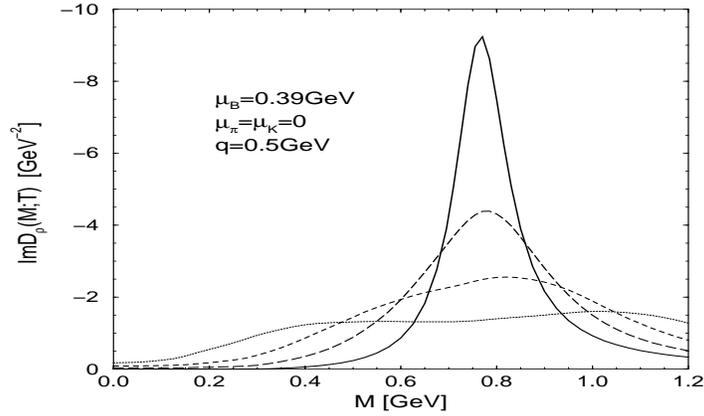,width=9cm,height=5cm,angle=-90}
\end{center}
\caption{\small{Imaginary part of the $\rho$ propagator in a hot 
hadron gas at temperatures $T$=127~MeV (long-dashed curve), $T$=149~MeV 
(dashed curve) and $T$=170~MeV (dotted curve) as well as in vacuum 
(full curve).} }
\end{figure} \\
For calculating $e^+e^-$ invariant mass spectra as measured in the
CERES experiment the  differential rate eq.~(\ref{rate}) has to be
integrated over 3-momentum and the space-time history of a
central 200~GeV/u S-Au reaction. For that we assume a temperature/density 
evolution as found in recent transport calculations~\cite{LKB}. The  
experimental acceptance cuts on the dilepton tracks as well as the 
finite mass resolution of the CERES detector are also included. 
We supplement our results for $\pi^+\pi^-$ annihilation with 
contributions from free Dalitz decays ($\pi_0$,$\eta$~$\to$~$\gamma e^+e^-$,
$\omega$~$\to$~$\pi^0e^+e^-$) and free $\omega$~$\to$~$e^+e^-$ decays
as extracted from ref.~\cite{LKB}, where medium effects are expected to be of 
minor importance. 
As can be seen from the  fig.~3, the use of the in-medium $\rho$ propagator 
(full curve) leads to reasonable agreement with the experimental $e^+e^-$ 
spectrum as observed in central S+Au collisions at 200~GeV/u. The same is true 
when comparing to the preliminary data for the heavier 
Pb+Au system at 158~GeV/u  
(see fig.~4), where an accordingly modified temperature/density evolution has 
been employed.  
\begin{figure}[!t]
\begin{center}
\epsfig{file=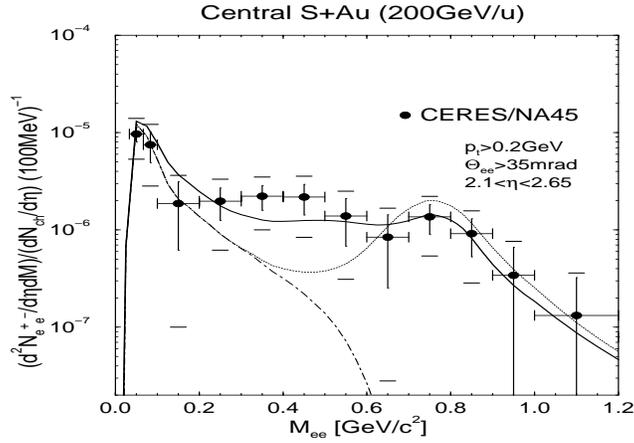,width=9cm,height=5cm,angle=-90}
\end{center}
\caption{\small{Dilepton spectra in central S+Au collisions from free 
Dalitz decays (dashed-dotted line), free Dalitz+$\omega$+$\rho$ 
(dotted line) and free Dalitz+$\omega$ + in-medium $\rho$ decays (full line).} }
\end{figure}
 
\begin{figure}[!b]
\begin{center}
\epsfig{file=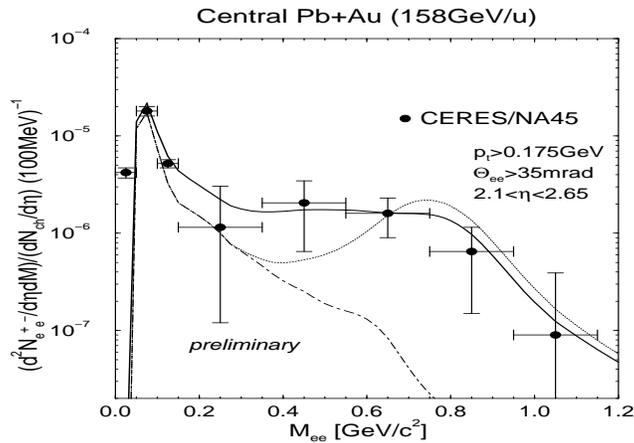,width=9cm,height=5cm,angle=-90}
\end{center}
\caption{\small{Same as fig.~3, but for central Pb+Au collisions.  } }
\end{figure}

To summarize, our findings seem to indicate that hadronic rescattering 
processes in in-medium $\rho$ propagation seem to resolve the discrepancy 
between the experimentally observed $e^+e^-$ enhancement at CERN-SpS energies 
and theoretical results based on free $\pi^+\pi^-$ annihilation. 
Even though further improvements to our analysis need to be done, we tend 
to conclude that the BR-scaling conjecture of a dropping $\rho$ mass is 
presumably not an independent phenomenon. For disentangling such a uniform 
mass shift from the dynamic mechanisms we discussed, the measurement of 
invariant mass spectra in various $p_T$ bins might provide new insights.  

\vskip0.5cm

\centerline {\bf ACKNOWLEDGMENTS}
Useful discussions with B. Friman and  A. Drees are gratefully 
acknowledged.
One of us (RR) acknowledges financial support from the
Alexander-von-Humboldt foundation as a Feodor-Lynen fellow.
This work is supported in part by the National Science Foundation
under Grant No. NSF PHY94-21309 and by the U.S. Department of Energy
under Grant No. DE-FG02-88ER40388.

\end{document}